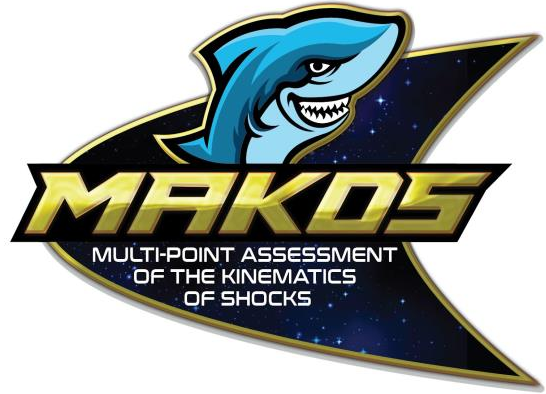

# Multi-point Assessment of the Kinematics of Shocks (MAKOS): A Heliophysics Mission Concept Study

K. A. Goodrich[1], L. B. Wilson III[2], S. Schwartz[3], I. J. Cohen[4], D. L. Turner[4], P. Whittlesey[5], A. Caspi[6], R. Rose[6], K. Smith[6]

*Endorsed by:*
R. Allen[4], D. Burgess[7], D. Caprioli[8], P. Cassak[1], J. Eastwood[9], J. Giacalone[10], I. Gingell[11], C. Haggerty[12], J. Halekas[13], G. Hospodarsky[13], G. Howes[13], J. Juno[14], Y. Khotyaintsev[15], K. Klein[10], H. Kucharek[16], B. Lembège[17], E. Lichko[10], T. Liu[18], D. Malaspina[3], M. F. Marcucci[19], C. Mazelle[20], K. Meziane[21], F. Plaschke[22], A. Retino[23], C. T. Russell[18], E. Scime[1], D. Sibeck[2], M. Stevens[24], J. TenBarge[25], I. Vasko[5], L. Wang[26], S. Wang[27], H. Zhang[28]

[1]West Virginia University, [2]NASA Goddard Space Flight Center, [3]University of Colorado Boulder, [4]The Johns Hopkins University Applied Physics Laboratory, [5]University of California Berkeley, [6]Southwest Research Institute, [7]Queen Mary University of London, [8]University of Chicago, [9]Imperial College London, [10]University of Arizona [11]University of Southampton, [12]University of Hawaii, [13]University of Iowa, [14]Princeton Plasma Physics Laboratory, [15]Swedish Institute of Space Physics, [16]University of New Hampshire, [17]LATMOS - CNRS- IPSL-UVSQ, [185]University of California Los Angeles, [19]INAF, [20]IRAP CNRS University of Toulouse CNES, [21]University of New Brunswick, [22]IGEP, TU Braunschweig, [23]LPP, CNRS, Sorbonne Université, Université Paris Saclay, Observatoire de Paris, Ecole Polytechnique Institut Polytechnique de Paris, [24]Smithsonian Astrophysical Observatory, [25]Princeton University, [26]University of Maryland at College Park, [27]Peking University, [28]University of Alaska Fairbanks

**Synopsis:**

Collisionless shocks are fundamental processes that are ubiquitous in space plasma physics throughout the Heliosphere and most astrophysical environments. Earth's bow shock and interplanetary shocks at 1 AU offer the most readily accessible opportunities to advance our understanding of the nature of collisionless shocks via fully-instrumented, in situ observations. One major outstanding question pertains to the energy budget of collisionless shocks, particularly how exactly collisionless shocks convert incident kinetic bulk flow energy into thermalization (heating), suprathermal particle acceleration, and a variety of plasma waves, including nonlinear structures. Furthermore, it remains unknown how those energy conversion processes change for different shock orientations (e.g., quasi-parallel vs. quasi-perpendicular) and driving conditions (upstream Alfvenic and fast Mach numbers, plasma beta, etc.). Required to address these questions are multipoint observations enabling direct measurement of the necessary plasmas, energetic particles, and electric and magnetic fields and waves, all simultaneously from upstream, downstream, and at the shock transition layer with observatory separations at ion to magnetohydrodynamic (MHD) scales. Such a configuration of spacecraft with specifically-designed instruments has never been available, and this white paper describes a conceptual mission design – MAKOS – to address these outstanding questions and advance our knowledge of the nature of collisionless shocks.

## 1. Scientific Motivation

Collisionless shocks are a fundamental plasma process. In astrophysical plasmas, shocks are responsible for converting kinetic bulk flow energy into plasma heat, enthalpy, plus nonthermal features, acceleration of suprathermal particles, and the excitation of a variety of linear to nonlinear plasma waves and kinetic structures. Understanding collisionless shocks is vital to the understanding of our plasma universe, from the heating and deflection of bulk flows to the acceleration of cosmic rays. Moreover, collisionless shocks directly influence our own terrestrial space environment, e.g., the bow shock's role in solar wind – magnetosphere interactions and space weather in Earths' magnetosphere-ionosphere-thermosphere system.

Despite that importance, plus decades of observations and theoretical/simulation studies, the basic ability to predict how a shock with given upstream parameters will partition the incident energy amongst the various degrees of freedom available remains elusive. Moreover, the various kinetic processes that perform that energy conversion within the shock remain impossible to resolve or only partially resolvable in prior and current observations. Goodrich et al.[1] lays out the questions that need to be answered to enlighten us, including the reasons why existing and previous missions and datasets cannot provide a complete answer.

The heliophysics community recognizes the importance of fundamental processes through the support of the previously-launched Magnetospheric Multiscale (MMS) mission (studying the fundamental process of magnetic reconnection) and the recently-selected Helioswarm mission (studying the fundamental process of turbulence). In order to achieve a complete view of the fundamental physics that dominate our universe, collisionless shocks must also be considered a subject of the highest significance in heliophysics. This can and must be done by supporting targeted and focused opportunities to observe the terrestrial bow shock in-situ, as detailed in the mission concept for the **Multipoint Assessment of the Kinematics of Shocks** (**MAKOS**) mission.

We propose the Decadal Survey Steering Committee consider highlighting the implementation of the **MAKOS** mission, which is specifically designed to study a significant parameter range of collisionless shocks. The MAKOS mission is a four-satellite mission designed to target the terrestrial bow shock, with the capabilities of also observing interplanetary shocks. MAKOS must be executed to address critical questions surrounding collisionless shocks. These questions are:

1) *What is the partition of energy across collisionless shocks?*
2) *What are the processes governing energy conversion at and within collisionless shocks?*
3) *How and why do these processes vary with macroscopic shock parameters?*

Exhibit 1 shows an abridged version of the MAKOS Science Traceability Matrix with the mapping to the science questions listed above. The following sections summarize the MAKOS mission design, instrumentation suite, cost/risk analysis, and enhancing technological developments.

## 2. Investigation Description

### *2.1 Mission Overview*

The baseline MAKOS mission concept (CML 4) comprises four spacecraft (S/C) with varying spatial separations at ion-kinetic to MHD scales in high-altitude, slightly elliptical (23.1×18.0 $R_E$) five-to-one (5:1) lunar resonance orbits (LROs) with oppositely oriented lines of apsides that maximize the number of bow shock crossings, even when apogee is on the nightside. Each of the two orbits has two S/C with separations on the order of ~100 to ~1000 km to obtain the required simultaneous upstream, downstream, and transition layer observations at shocks, including multipoint observations at ion-kinetic scales through every shock transition layer crossing. The separations between the S/C on the different orbits range from ~5 to 12 $R_E$. This provides year-round crossings of the bow shock with simultaneous multipoint separations ranging from ion kinetic (100~1000 km; each pair) to MHD (several $R_E$; the pair of pairs) scales, as well as prolonged dwell time throughout the year in the solar wind, enabling opportunities to also study interplanetary (IP) shocks and for MAKOS to simultaneously probe electron- and ion-kinetic plus MHD-scale processes during every single bow shock and IP shock crossing (>1000 expected during MAKOS' 2-year prime mission).

| MAKOS Science Traceability Matrix | | | | | Instrument Requirements | | | |
|---|---|---|---|---|---|---|---|---|
| Science Questions | Science Objectives | Physical Parameters | Observable Quantities | Instrument | Instrument & Parameter | | Measurement Req. | Exp. Data Volume per Orbit |
| [Q1] What is the energy budget both upstream and downstream of a collisionless shock? | Quantify the contribution of proton and electron thermal and kinetic energy to the shock energy budget | Simultaneous upstream and downstream moments (density, velocity, pressure, heat flux) of particle sub-populations | Simultaneous upstream and downstream core 3D velocity distribution functions | SWI | SWI | Energy Range Energy Resolution FOV Angular Resolution Temporal Resolution | 300 eV – 7 keV 10% 40° x 40° 6° 0.1 s | 27 GB |
| | | | | SWE | SWE | Energy Range Energy Resolution Angular Coverage Angular Resolution Temporal Resolution | 3 eV – 1.5 keV 10% 4π-sr 20° 0.01 s | 314 GB |
| | Quantify the contribution of He and the CNO group thermal and kinetic energy to the shock energy budget | | Simultaneous upstream and downstream suprathermal 3D velocity distribution functions | STI | STI | Energy Range Energy Resolution Angular Coverage Angular Resolution Temporal Resolution | 700 eV – 30 keV 20% 4π-sr 20° 1 s | 102 GB |
| | | | | STE | | | | |
| | | | Simultaneous upstream and downstream energetic particle energy, angular, and compositional distributions | EP | STE | Energy Range Energy Resolution Angular Coverage Angular Resolution Temporal Resolution | 500 eV – 30 keV 20% 4π-ster 20° 1 s | 13 GB |
| | Quantify the contribution of Poynting flux to the shock energy budget | Electric and Magnetic field contribution to the Poynting flux | Simultaneous upstream and downstream 3D DC- and AC-coupled electric and magnetic field | EFI | | | | |
| | | | | FGM | | | | |
| | | | | SCM | | | | |
| [Q2] What are the processes governing energy conversion at and within collisionless shocks? | Characterize the coherent and incoherent heating and acceleration of particle populations upstream, downstream, and within the shock front | Particle heating | Simultaneous upstream, within shock, and downstream core, suprathermal and energetic particle 3D Velocity Distribution Functions (VDFs) | SWI | EP | Energy Range Energy Resolution Species FOV Angular Resolution Temporal Resolution | 20 keV– 10 MeV 20% H, He, C, O, Ne, e- 180° 30° 1 s | 30 GB |
| | | | | SWE | | | | |
| | | | | STI | | | | |
| | | | | STE | | | | |
| | | | | EP | | | | |
| | Identify electric and magnetic field variations together with targeted local plasma instabilities and resulting waves within the shock | Non-Maxwellian features responsible for observed instabilities | Simultaneous upstream, within shock, and downstream core 3D VDFs | SWI | FGM | (DC) Dynamic Range Resolution Temporal Resolution | ±500 nT 10 pT 0.03125 s | 278 MB |
| | | | | SWE | | | | |
| | | | Simultaneous upstream, within shock, and downstream suprathermal 3D VDFs | STI | SCM | (AC) Dynamic Range Resolution Temporal Resolution | ±50 nT 0.1 pT 0.001 s | 12 GB |
| | | | | STE | | | | |
| | | Magnetic and electric field topology and wave modes | Simultaneous upstream, within shock, and downstream 3D DC- and AC-coupled magnetic and electric field | FGM | | | | |
| | | | | SCM | | | | |
| | | | | EFI | EFI | (DC) Range Dimensions Resolution Temporal Resolution | ±1000 mV/m 3 1 mV/m 0.5 s | 115 GB |
| [Q3] How and why do these processes vary with shock orientation and driving conditions? | Parameterize shock crossings according to the macroscopic, Rankine-Hugoniot relations | Particle-dependent macroscopic shock parameters. | Upstream, within shock, and downstream particle moments and 3D DC-coupled magnetic field | All | | | | |
| | Tabulate and sort observed shock crossings according to the macroscopic shock parameters for statistical analysis of science objectives | | Statistical parameterization of the processes in [Q1] & [Q2] versus calculated shock parameters | All | | (AC) Range Dimensions Resolution Temporal Resolution | ±2000 mV/m 3 1 mV/m 0.001 s | |

**Exhibit 1.** MAKOS will address outstanding questions regarding the cross-scale physical processes at play at collisionless shocks.

MAKOS requires each S/C to carry a comprehensive science payload of particles, fields, and waves instruments specifically tailored to measure the in-situ processes at play in collisionless shocks. The need to resolve microphysical phenomena in and around each shock and to fully characterize the plasma populations upstream and downstream of each shock drives a mission requirement that the complete three-dimensional thermal and suprathermal electron and ion velocity distributions be sampled at very high temporal resolution (<1 s). This is achieved in the notional mission design by carrying multiple dedicated sensors targeting each species and energy range on a rapidly-spinning (10 RPM baseline) S/C. Furthermore, the need to resolve the evolution of the solar wind ion beam necessitates a dedicated detector that is constantly pointed into the solar wind - i.e., along a S/C spin vector anti-aligned with the solar wind flow direction.

*2.2 Science Payload*

The resource demands of the MAKOS science payload are summarized in Exhibit 2.

**Solar Wind Ions (SWI).** Two SWI sensor heads - based on the Parker Solar Probe/SWEAP/SPAN-I instrument[2]-[3] - will be oriented such that their fan-like, planar (40°×~6°) fields-of-view (FOVs) are orthogonal to each other and both parallel to the nominal solar wind direction, i.e., roughly parallel to the S/C spin axis.

| Instrument | # | CBE per unit (kg) | CBE Total (kg) | CBE per unit (W) | CBE Total (W) |
|---|---|---|---|---|---|
| SWI | 2 | 3.5 | 7.0 | 3.5 | 7.0 |
| SWE | 4 | 2.6 | 10.4 | 3.2 | 12.8 |
| STI | 4 | 11.4 | 45.6 | 12.0 | 48.0 |
| STE | 4 | 2.6 | 10.4 | 3.2 | 12.8 |
| EP | 1 | 3.9 | 3.9 | 3.8 | 3.8 |
| FGM | 2 | 0.7 | 1.4 | 4.0 | 8.0 |
| SCM | 1 | 0.8 | 0.8 | 1.0 | 1.0 |
| EF | 1 | 22.0 | 22.0 | 8.4 | 8.4 |
| Totals | | | 101.5 | | 101.8 |

**Exhibit 2.** MAKOS carries a high-TRL (≥6) payload specifically tailored to measure the electromagnetic fields and particle populations required to understand energy partitioning and conversion processes at collisionless shocks.

**Solar Wind Electrons (SWE).** Four SWE detector heads - based on the WIND/3DP/EESA-L sensor[4] - will each view the sky with a fan-like >180°×3° FOV (coplanar with S/C spin axis) pointing outward at ~90° spacing around the S/C.

**Suprathermal Ions (STI).** Four STI detector heads - based on the STEREO/PLASTIC instrument[5] - will each view the sky with a fan-like ~180°×6° FOV (coplanar with the S/C spin axis) pointing radially outward at ~90° spacing around the S/C to achieve the 4π-sr sky coverage and temporal resolution required for MAKOS.

**Suprathermal Electrons (STE).** Four STE detector heads - based on the Wind/3DP/EESA-H instrument[4] - will each view the sky with a fan-like ~180°×14° FOV (coplanar with the S/C spin axis) – i.e., only half the EESA-H azimuthal range - pointing radially outward at ~90° spacing around the S/C to achieve the 4π-sr sky coverage and temporal resolution required for MAKOS.

**Energetic Particles (EP).** Each MAKOS S/C will carry a single EP sensor - based on the time-of-flight-by-total energy PSP/ISΘIS/EPI-Lo instrument[6]-[7] - with a nearly 2π-sr FOV.

**Fluxgate Magnetometer (FGM).** Two FGM sensors - based on the MMS/FIELDS/FGM tri-axial (orthogonal to within ~1°), fluxgate instrument[8]8-[9] - will be mounted on a common 5-m, single-hinged boom in a "gradiometer" configuration to characterize and eliminate S/C signals of electromagnetic interference. It is assumed that the main MAKOS/FGM electronics will be housed in a common "fields" electronics box housing along with those of the SCM and EF instruments.

**Search Coil Magnetometer (SCM).** The three-axis SCM sensor - based on three orthogonal (to within ~1°) instances of the search coil magnetometer of the Juno/WAVES instrument[10] - will be mounted on a second 5-m, single-hinged boom (identical but oppositely mounted from the FGM boom). The MAKOS/SCM electronics will also be housed in the common "fields" electronics box.

**Electric Fields (EF).** The three-axis EF instrument - based on the MMS/FIELDS/ADP (axial) and SPD (spin-plane) instruments[9][11][12] - will comprise twelve spherical voltage probes mounted on four 50-m wire booms in the spin-plane of the S/C and two 10-m stacer booms along its spin-axis (i.e., axial). EF will employ two probes (as implemented on the FAST mission[13]) - separated by ten meters - on each boom to accurately resolve wave phenomena with wavelengths 100 m.

### 2.3 MAKOS Spacecraft Reference Design

The MAKOS observatories comprise the MAKOS payload (Section 2.2) integrated with a spin-stabilized S/C using a single-string hardware architecture with functional and selective redundancy included for critical areas. The architectural approach achieves a mission success of >90% over its two-year mission lifetime as demonstrated by the NASA CYGNSS eight-observatory mission[14], which has operated for over five years without failure.

The simple operational nature of the MAKOS instruments and science profile allows significant autonomous on-board control of the observatory during all normal science and communication operations without need for daily on-board command sequences. Observatory initialization and science operations use five sub-modes: rate damping, nutation damping, Sun acquisition/precession control, spin-rate control, and science. After initial damping of launch

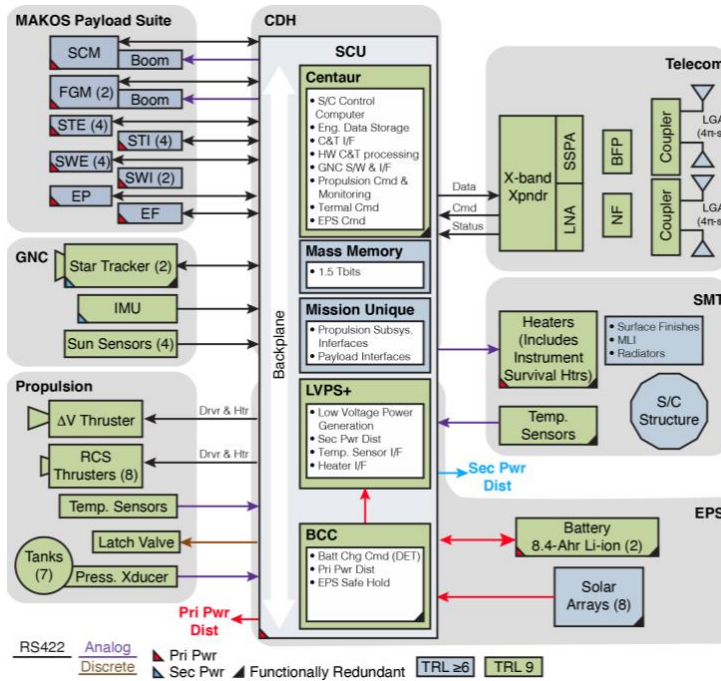

**S/C Functional and Selective Redundancy**

- Dual fault tolerant separation sense for initial power on
- Battery cell bypass diodes
- DET channels mapped to S/A strings allows loss of S/A string
- Backup H/W and S/W timers to ensure transmitters are powered off after communication passes have completed
- Heater and temperature sensor overlay to enable functional redundancy
- Propulsion functional redundancy
- Dual star trackers
- $4\pi$-sr Sun sensors (Safe Mode)
- 3 copies of S/C FSW are stored in MRAM and use cyclic boot tree that terminates into a "gold copy" FSW stored in write-protected boot storage
- H/W only "L0" command & telemetry capability for S/C recover if FSW fails

**Exhibit 3.** The MAKOS observatory reference design uses a single-string architecture with functional redundancy with heritage from the NASA CYGNSS mission.

vehicle separation rates is complete, the observatory transitions to Sun acquisition using Sun sensors and a sky-searching algorithm to locate the Sun vector. The observatory then uses reaction control thrusters to point the S/C solar arrays at the Sun using the rate and Sun sensors. The star trackers are initialized followed by spin-up of the S/C to its operational spin rate of 10 rpm. The vehicle spin axis then precesses to align with the local solar wind vector for science operations.

The MAKOS observatory design (Exhibit 3) is mission-specific to meet science requirements and instrument accommodations. Physical accommodation of the MAKOS instruments and spin stabilization implementation drives the observatory's structure and thermal design. Fixed solar arrays, located on the Sun-oriented face of the observatory provide electrical power for the S/C. The LRO enables use of a simple direct energy transfer architecture for battery charging with the batteries sized to accommodate full science operations during solar eclipse periods. Primary attitude knowledge is star tracker-based augmented with rate sensors for stability and nutation determination. Sun sensors are included for emergency operations. Observatory orientation, spin-rate and precession are all controlled using an on-board cold-gas $SF_6$-based reaction control subsystem. Observatory positional knowledge is based on GPS receivers augmented by on-board optical navigation during GPS outages. Communication is provided by an X-band transponder and low-gain patch antennas to provide communications without interrupting science operations. On-board timing requirements are driven by science data synchronization within the constellation relative to measurement of the solar wind and electric field waveforms. Specific S/C performance characteristics are provided in Exhibit 4. Observatory magnetic and electrostatic cleanliness is key to the MAKOS instruments meeting science Level 1 requirements. MAKOS uses mature electromagnetic requirements consistent with previous missions (e.g., MMS, Cluster, THEMIS) to develop a magnetically and electrostatically clean observatory.

| Parameter/Item | | Performance | Parameter/Item | | Performance |
|---|---|---|---|---|---|
| Structure | Type | CFRP panels with milled Al supports | Space to Ground Communications | Uplink | X-band 256 kbps |
| | Size | Octagonal; 2.0m x 0.65m | | Data Downlink | Science: 4 Mbps Engineering: 256 kbps |
| | 1st Mode | >210 Hz | | | |
| | SV Mass (dry) | 315.7 kg | Data | Data Storage | 188 GBytes |
| Thermal | Architecture | Cold bias | Attitude Knowledge | Star Tracker | Dual, 20 arc-sec (1σ) |
| | Control | Heaters, MLI, radiators | | IMU | Bias Stability: 0.3°/hr |
| Solar Array | Configuration | 6-panel, body mounted | | Performance | <30 arc-sec (1σ) |
| | Size | 0.84m² | Attitude Control | Architecture | Spin stabilized, 10 rpm |
| | Cell Type | Triple junction with AR coating | | Pointing | <1.3 deg |
| | Cell Efficiency | 28.4% (EOL) | | Nutation | <1.6 deg (p-p) |
| | Full power output | 283 W (EOL) | Orbital Knowledge | GPS position | <100 m (1σ) |
| Battery | Configuration | 8 cells/10 strings (8p10s) (x2) | | Velocity | <10 cm/s (1σ) |
| | Cell type | Li-ion | Propulsion | Type | Cold Gas (SF$_6$) |
| | Capacity | 56 Ahr | | DeltaV Thrusters | 1N, Isp: 45 sec (qty 3) |
| | DOD during full eclipse periods | <49% with no operational restrictions | | RCS Thrusters | 120mN, Isp: 45 sec (qty 8) |
| Power | Average Load | 149.5 W (cold case) | | Delta-V | >160 m/s when fully loaded |
| | Margin | 90% | | Reaction Control | 6 degrees of freedom |

**Exhibit 4.** The MAKOS S/C performance is more than sufficient to meet the mission's science requirements and instrument accommodations.

Expected science data generated is 53.5 GB/orbit. Baseline on-board data storage provides 188 GB or 3.5 orbits of science data storage to allow for recovery from downlink anomalies. The baseline reference communication uses a 14 Mbps X-band RF link that, with 14% overhead for CCSDS, requires ~9.6 hr to downlink science data from 1 orbit. Significantly improved data rates would be available to reduce downlink durations and/or increase data downlink quantities if the optical communications are realized prior to MAKOS implementation (see Section 4.2).

## 2.4 Concept of Operations

Telemetry is a major driver of the notional MAKOS mission design as the science requires very high data rates for observatory science telemetry at and around each collisionless shock crossing. Furthermore, MAKOS should also capture the highest rate data from any interplanetary shocks encountered upstream of the bow shock. Despite this, the MAKOS concept of operations (CONOPS) (Exhibit 5) is simple by design and consists of collecting science data (telemetry) from each of the four identical observatories during the two-year prime science mission. Each observatory will record telemetry in one of two science modes: i) high-rate and ii) low-rate. Even under extreme solar wind driving conditions, the bow shock is consistently located outside of the *average* (i.e., typical) magnetopause location. Thus, the average magnetopause location offers an opportune surface to use for routine orbit-to-orbit operations and systematically toggling the MAKOS S/C between high- (i.e., along the orbit *beyond* the average magnetopause location) and low-rate (i.e., along the orbit *within* the average magnetopause location) modes. Using the average magnetopause location and the orbit predicts to schedule onboard science telemetry mode changes, each MAKOS observatory shall switch from low- to high-rate data collection when it transits from the magnetopause into the magnetosheath (i.e., outbound model magnetopause crossings), and each observatory will switch from high- to low-rate data collection when it transits from the magnetosheath into the magnetosphere (i.e., inbound magnetopause crossings).

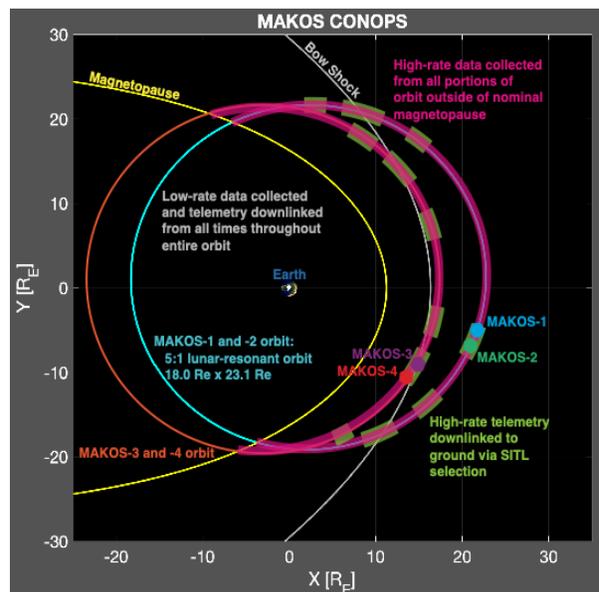

**Exhibit 5.** MAKOS uses two identical, 180°-phased LROs to achieve its target inter-S/C separations. A proven SITL process will be used to prioritize high-rate data obtained during predefined portions of the orbit for downlink.

The MAKOS payload generates data at either 807 kbps (low-rate) or 20.875 Mbps (high-rate) to achieve the temporal resolutions required for each observable. Acquiring high-rate data only when the S/C are *sunward* of the average magnetopause requires high-rate telemetry being recorded for ~60 hrs (46%) of each 5.46-day orbit.

However, all 614 GB of science data cannot be transmitted to ground each orbit because of limitations of the communications subsystem and ground network. To ensure that all collisionless shock transits are captured during the prime mission, MAKOS will employ a "scientist-in-the-loop" (SITL) strategy similar to that used by MMS[15]. A trained MAKOS science and data expert (i.e., SITL) will review a special low-rate data product produced onboard and telemetered to ground each orbit to make prioritized selections of which periods of the high- and low-rate data shall be telemetered to the ground. Shock crossings will be prioritized, and data from and around each shock crossing will be telemetered to the ground to ensure prime science closure. The expected SITL-selected high-rate data volume averages 5.9 GB (~1% of recorded data) per S/C per orbit; combined with the low-rate data generated each orbit (47.6 GB per S/C), this yields 53.5 GB of data to be telemetered to ground from each MAKOS S/C each orbit (9% of 595 GB total recorded data). Over the two-year prime mission, all four MAKOS S/C will telemeter 28.6 TB of total 315 TB scientific data recorded.

| MAKOS Concept Study | FY22 $M |
|---|---|
| *Phase A* | *Not incl.* |
| Project Management | 39.8 |
| Systems Engineering | 23.4 |
| Safety & Mission Assurance | 22.9 |
| Science / Technology | 77.4 |
| Instruments | 224.1 |
|    Solar Wind Ions | 15.5 |
|    Solar Wind Electrons | 22.8 |
|    Suprathermal Ions | 81.3 |
|    Suprathermal Electrons | 35.3 |
|    Energetic Particles | 13.0 |
|    Fluxgate Magnetometer | 6.5 |
|    Search Coil Magnetometer | 4.4 |
|    Electric Fields | 29.9 |
|    Payload Electronics | 15.3 |
| Spacecraft | 143.4 |
| Mission Operations | 46.4 |
| Launch Vehicles | 0.0 |
| Ground Systems | 14.1 |
| Observatory Integration & Test | 59.5 |
| Subtotal before reserves | 650.9 |
| Reserves @ 50% B-D, 25% E-F | 312.9 |
| **Total** (excl. Phase A & Launch Services) | **963.8** |

**Exhibit 6.** MAKOS cost by WBS.

## 3 Mission Cost, Risk, and Schedule

Exhibit 6 presents a baseline cost summary estimated using multiple parametric model sets in parallel. The four-observatory configuration will require $651M (FY22) funding as a current best estimate. Recognizing that this is a preliminary concept study, conservative reserves are applied to all cost elements: 50% for all Phase B-D work and 25% for Phase E-F. This brings the baseline estimate to $964M (with NASA's addition of a Phase A study and Launch Services to complete the funding). Additional development costs are not included, as baseline instruments and supporting hardware were chosen to be at TRL 6 prior to Phase A. Independent analysis by the NASA Goddard Cost Estimating, Modeling and Analysis (CEMA) Office concluded that the MAKOS baseline budget with proposed reserves is more than adequate to fund this project.

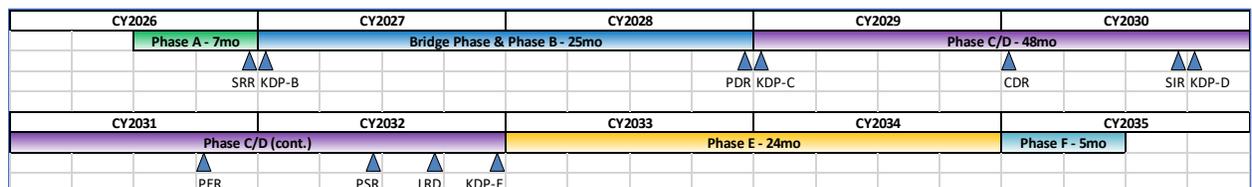

**Exhibit 7.** The MAKOS development plan baselines 79 months for Phases A–D, with a staggered observatory AI&T approach - including a 3-month cruise and commissioning, followed by a 2-year Phase E.

| # | Risk | Type | L | C | Mitigation |
|---|------|------|---|---|------------|
| 1 | IF a launch issue precludes all four S/C from achieving the necessary formation, THEN there could be delay to the science phase and/or impact to science closure. | Cost, Schedule | 1 | 5 | Phase A trades will consider additional propulsion capacity in S/C design to potentially enable achievement of baseline MAKOS configuration from a single launch. |
| 2 | IF instrument cross-calibration requires more analysis to resolve known challenges and ensure data product adequacy, THEN additional effort would be required. | Cost, Technical | 3 | 2 | Use of advanced data analytic techniques to develop novel ways to cross-correlate the data using timing, position, and events to improve completeness of datasets for science would be required. |
| 3 | IF specialized component updates are needed for the EF instrument deployment mechanism, THEN additional development effort would be required. | Cost, Technical | 2 | 2 | Additional design, prototyping, and testing will be conducted to reduce likelihood of failure of the EF deployment mechanism. |

**Exhibit 8.** MAKOS has no severe mission risks and favorable mitigations for the top three identified risks. (L= risk likelihood, C= risk consequence)

The MAKOS risk assessment combines with cost to identify key risks likely to drive significant variances if not managed. A four-observatory constellation, each carrying eight instruments, is within the overall experience base of the institutional partners, but the need to deliver multiple flight units raises the criticality of certain common development issues. Exhibit 7 shows the top-level schedule with major milestones. Exhibit 8 lists the top three identified risks and potential mitigations.

## 4 Enhancing Technology Development Needs

**Instrument Development.** Obtaining more comprehensive 3D particle measurements at cadences even faster (e.g., 10-ms) than recent missions (e.g., MMS and Parker Solar Probe) – without relying on a high number of sensors - will require additional instrument development for traditional top-hat ESAs or development of new particle detection systems for low-energy space plasmas. Particular emphasis is needed in two key areas: 1) parts availability, e.g., reliable high voltage optocouplers, and 2) tuning and responsiveness of the high voltage power supplies to ensure fast measurements are being taken with sufficient accuracy. At least one vendor that has provided flight parts for previous NASA missions has existing custom optocoupler designs that can fulfill even the most ambitious high-resolution MAKOS measurement cadences.

**Infrastructure.** While MAKOS achieves its baseline science with current RF communications infrastructure, it requires limiting high-rate data collection to only targeted portions of the orbit. Even downlinking data only when S/C are earthward of the magnetopause (i.e., ~71-hr/orbit window) requires hours per day per S/C of Deep Space Network (DSN) time. Optical communications would drastically reduce required downlink, thus enabling significantly more science data to be downlinked and reducing SITL decisions and complexity. The much higher data rates afforded by optical downlink would enhance MAKOS by significantly reducing resource competition and/or providing additional science data and reducing the need for SITL-based operations.

## 5 Conclusion

MAKOS is an exciting new multi-spacecraft mission – the first ever with a comprehensive payload specifically designed to address outstanding questions and advance our knowledge of the nature of collisionless shocks. Specifically, MAKOS will provide novel multipoint observations that will enable direct measurement of the necessary plasmas, energetic particles, and electric and magnetic fields and waves, all simultaneously from upstream, downstream, and at the shock transition layer with observatory separations at ion to magnetohydrodynamic (MHD) scales.